\documentclass[aps,onecolumn]{revtex4}

\usepackage{amsfonts}
\usepackage{amsmath}
\usepackage{amssymb}
\usepackage{version}
\usepackage{graphicx}

\includeversion{New_connection}
\excludeversion{Old_connection}
\newtheorem{theorem}{Theorem}
\newtheorem{acknowledgement}[theorem]{Acknowledgement}

\begin{document}

\preprint{}
\title{A Reassessment of the Evidence for Macroscopic Quantum Tunneling in a
Josephson Junction}
\author{James A. Blackburn}
\affiliation{Department of Physics and Computer Science\\
Wilfrid Laurier University\\
Waterloo, Ontario, Canada}

\begin{abstract}
Switching current distributions have for decades been an indispensable
diagnostic tool for studying Josephson junctions. \ They have played a key
role in testing the conjecture of a macroscopic quantum state in junctions
at millikelvin temperatures. \ The conventional basis of the test has been
the observation of temperature independence of SCD\ peak widths, and that
led to affirmative conclusions about a crossover. \ A different criterion is
proposed here - the distance of the SCD peak from the junction critical
current - and its efficacy is demonstrated. This test has distinct
advantages in terms of precision, and it is found that, for three example
experiments, the evidence for a crossover to the conjectured macroscopic
quantum state is unequivocally negative. \ The implications of this finding
for superconducting qubits are considered.
\end{abstract}

\maketitle

\section{Introduction}

Soon after the Josephson effect had been confirmed in 1963 a classical
equivalent circuit for a Josephson junction was proposed, the so-called RCSJ
model. For many decades this RCSJ model was successful in predicting the
outcome of experiments on superconducting circuits that employed Josephson
devices, and it remains so. \ At the beginning of the 1980s a hypothesis was
advanced \cite{Leggett} which suggested that a Josephson junction, at
temperatures typically below about $100mK$, would crossover to a
\textquotedblleft macroscopic quantum\textquotedblright\ state. If true,
this was thought to give rise to a transformation of the junction into an
\textquotedblleft artificial atom\textquotedblright\ which was expected to
open the door to the development of Josephson qubits. Above about $100mK$
the RCSJ model would still be appropriate, but below that temperature the
classical RCSJ model would give way to quantum descriptions.

Experiments were devised for the purpose of confirming the existence of this
macroscopic quantum state by seeking evidence of macroscopic quantum
tunneling (MQT), just as the earlier experiments had confirmed the Josephson
effect itself. \ These new efforts were based on interpretations of the
distributions of switching currents that are exhibited in swept bias
experiments. As shown in \cite{BCJJAP}, the crossover, if it exists, would
have a distinct signature that could be easily detected. The first report of
success at seeing the crossover effect appeared in 1981 \cite{VossWebb}. \
This was followed by a number of studies, all seeming to corroborate the
quantum hypothesis. \ By the end of the 1980s, it was generally considered
that the macroscopic quantum state had been validated, and from then on a
junction operated at the base temperature of a dilution refrigerator could
be presumed to be an artificial atom.

Here the evidence for that presumption is reassessed. A new test is
introduced for determining when, or if, a Josephson junction exhibits
evidence of macroscopic quantum tunneling, and the test is applied to data
from three independent swept bias experiments dating from 1981 to 2013.
Theory and experiment are compared, especially in the crucial region below $
200mK$.

\section{Switching Current Distributions}

A Josephson junction with phase $\varphi $ has stored potential energy $%
E_{J}(1-\cos \varphi )$. The pre-factor in this expression is the Josephson
energy:

\begin{equation}
E_{J}=\hbar I_{C}/2e  \label{Eq3}
\end{equation}

The total potential energy of a junction, when an additional bias current $I$
is supplied, is

\begin{equation}
U=E_{J}\left\{ \left( 1-\cos \varphi \right) -\eta \varphi \right\}
\label{Eq4}
\end{equation}%
where $\eta =I/I_{C}$. \ This is the well known washboard potential which
consists of periodic wells due to the $(1-\cos \varphi )$ term superimposed
on the tilting action of the bias current. \ If the applied current is less
than the junction critical current ($\eta <1$) then the system resides in a
well of finite depth, but at $\eta =1$ the well has zero depth. \ The
junction can escape from the well via classical thermal activation \cite
{Kramers}, at a rate

\begin{equation}
\Gamma _{TA}=f_{J}\exp \left( \frac{-\Delta U}{k_{B}T}\right)
\label{GAMMATA}
\end{equation}%
where $f_{J}=f_{J0}\left( 1-\eta ^{2}\right) ^{1/4}$ with $f_{J0}$ being the
zero bias Josephson plasma frequency, $T$ is the temperature, and $\Delta U$
is the barrier height. 
\begin{equation}
\Delta U=2E_{J}\left[ \sqrt{1-\eta ^{2}}-\eta \cos ^{-1}\eta \right]
\label{Barrier}
\end{equation}%
\ The barrier height and the plasma frequency are both directly set by the
bias current $\eta $; as it changes, they change.

If the bias current is steadily increased from zero, then at some moment the
junction will escape from the potential well and for light damping bounce
down successive wells. \ This will be indicated by an abrupt switch to a
non-zero junction voltage. If the experiment is repeated, the bias at which
the junction voltage switches will be different. \ However if these sweeps
are repeated many times, at the same temperature, then the accumulated
switching current data may be plotted and a switching current distribution
(SCD) peak is the result.

\begin{figure}
[h]
\begin{center}
\includegraphics[
height=2.660in,
width=5.894in
]
{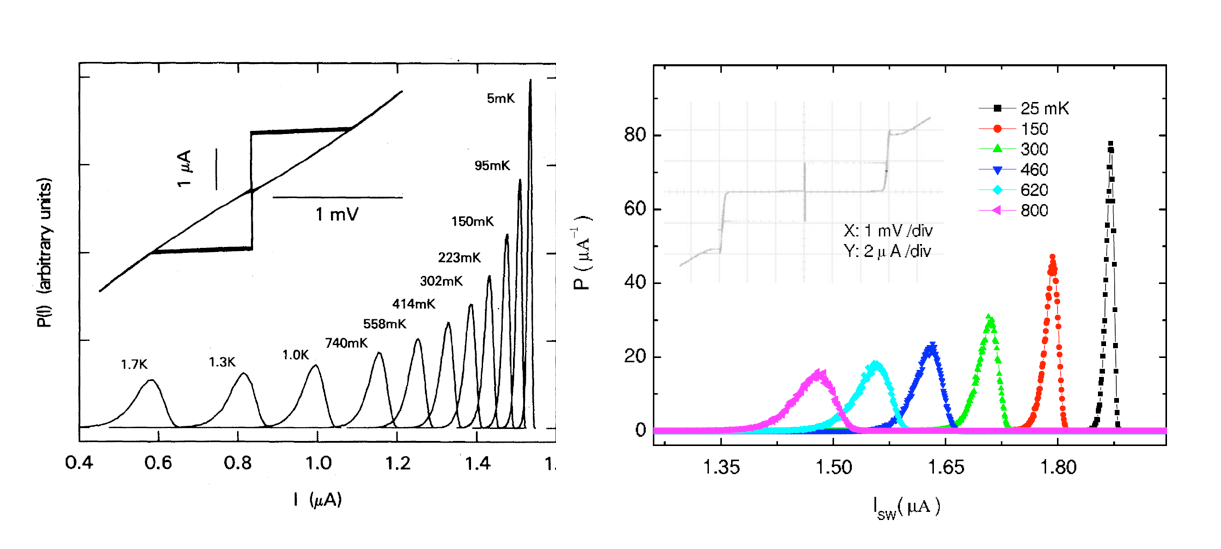}
\caption{Experimental switching current
distributions for a number of different temperatures. \ Left: Fig.1 from 
\protect\cite{VossWebb}; Right: Fig.1 from \protect\cite{Yu}.}
\label{SCDPeaks}
\end{center}
\end{figure}

These SCD peaks are the experimental data from which the behavior of the
Josephson junction can be surmised. \ Examples of SCD data from two
experiments are shown in Fig. \ref{SCDPeaks}. Generally speaking, the
behavior of switching current peaks at temperatures above about $150mK$ is
in accord with classical thermally induced escape processes.

Swept bias experiments have been central to searches for other phenomena
such as microwave induced transitions \cite{microwaves}.

\section{Macroscopic Quantum Tunneling\ }

\ In a macroscopic quantum state escape from a potential well to a finite
voltage running mode could happen via macroscopic quantum tunneling (MQT)
through the potential barrier, rather than\ by thermal activation (TA) over
the barrier.

The escape rate for this process is given by (see e.g. \cite{Devoret} \ \cite
{Martinis}).

\begin{equation}
\begin{array}{l}
{\Gamma_{MQT}=a_{q}\;f_{J}\exp\left[ -7.2\frac{\Delta U}{hf_{J}}\left( 1+%
\frac{0.87}{Q}\right) \right] } \\ 
{a_{q}\approx\left[ 120\pi\left( \frac{7.2\Delta U}{hf_{J}}\right) \right]
^{1/2}}%
\end{array}
\label{Eq18}
\end{equation}
where $Q$ is the quality factor of the junction ($Q=\omega_{J}RC$).

The expression for escape due to quantum tunneling, Eq.(\ref{Eq18}), is a
limiting form generally considered applicable only for the lowest
temperatures. It is a widely held opinion that any judgment as to whether
quantum theory actually does describe experiments must include temperature
enhancement effects in a revised expression for the escape rate. MQT escape
rates are expected to increase slightly with finite temperature. We note
that an enhanced escape rate (greater than $\Gamma _{MQT}$) means it is
easier to escape from the well, so escape will occur sooner in the sweep.
Therefore a finite temperature effect will result in SCD peaks being shifted
to lower bias positions. For this reason the MQT escape peak expected from
Eq.(\ref{Eq18}) must represent \ a \textit{maximum} possible bias position
for an SCD peak; no quantum peak can advance beyond this point no matter how
low the sample temperature. \ So there should be a \textquotedblleft
cutoff\textquotedblright\ value for activation peak positions.

This finite temperature effect, as it applies to the particular case of
Josephson junctions, appeared in \cite{Grabert} where the enhanced escape
rate in the weak damping limit was obtained from the zero temperature rate
Eq.(\ref{Eq18}) according to the following expressions (Eqs.3.16, 3.3, 3.11
in \cite{Grabert}):

\begin{equation}
\ln\left[ \Gamma(T)/\Gamma(0)\right] =10\pi\alpha(B-\frac{8}{5}%
)(k_{B}T/hf_{J})^{2}  \label{Martinis}
\end{equation}
where:

\begin{align}
B& =(\Delta U/\hbar \omega _{J})s(\alpha )  \label{parameter1} \\
s(\alpha )& =\frac{36}{5}\left[ 1+\frac{45}{\pi ^{3}}\xi (3)\alpha \right]
\label{parameter2}
\end{align}%
with a damping constant $\alpha =1/2Q$ for the Josephson junction and $\xi
(3)=1.202$ is a Riemann number.

The most obvious attribute of macroscopic quantum tunneling, Eq.(\ref{Eq18}
), is its temperature \emph{independence}. \ As was demonstrated in \cite
{BCJJAP}, even accounting for thermal enhancement, only a small local region
around the crossover point is affected, below which temperature independence
occurs. \ Therefore, temperature independence in SCD peaks\ is the \emph{%
hallmark} of a macroscopic quantum state in a Josephson junction.

\section{Prevailing Methods for Interpreting SCD Peaks}

In 1981 R.F. Voss and R.A. Webb \cite{VossWebb} reported \textquotedblleft
the first compelling evidence for the existence of quantum tunneling of a
macroscopic variable\textquotedblright\ in a Josephson junction. \ They
chose to present their experimental data in the form of a plot of SCD peak 
\emph{width} versus log temperature. \ The primacy given to peak width was
possibly due to the known $T^{2/3\text{ }}$dependence of width expected of
the thermal activation process. \ In other words, it might have been thought
that a log temperature scale would focus on any deviations from TA theory.\
However, the supposition that data which deviated from TA expectations would
necessarily fall ino a category of vindicating MQT thery is certainly flawed.

\begin{figure}
[h]
\begin{center}
\includegraphics[
height=4.331in,
width=5.5827in
]
{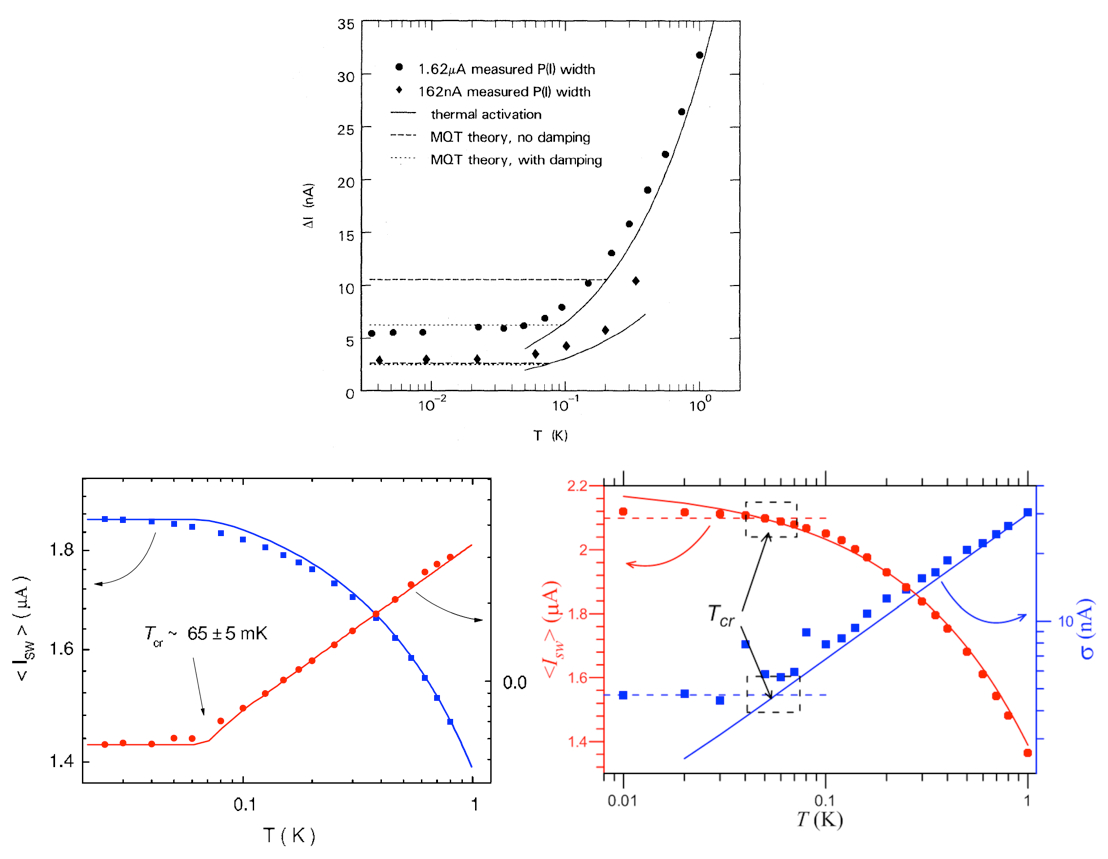}
\caption{Plots of SCD peak data from three
experiments. Upper panel: peaks widths versus temperature from Voss \& Webb 
\protect\cite{VossWebb}. \ Note that only the widths of the peaks were
presented. Lower panel, left: Results from Yu et al.\protect \cite{Yu}
displaying both peak positions (left vertical linear scale) and peak widths
(right vertical log scale). Lower panel,right: Results from Oelsner et al. 
\protect\cite{Oelsner} displaying both peak positions (left axis) and peak
widths (right vertical axis with log scale).}
\label{scdpeakwidths}
\end{center}
\end{figure}

The upper panel in Fig. \ref{scdpeakwidths} is Fig.3 from \cite{VossWebb}. \
The stretching effect of the log temperature scale is quite evident. \ The
dotted line labelled \textquotedblleft MQT theory with
damping\textquotedblright\ corresponds to a supposed crossover point at $%
T_{cr}=100mK$, \ This presentation is meant to foster the impression that
the width data switch to a freezing line, but a careful inspection shows
this is not the case - the peak widths continued to decline as the
temperature was lowered. \ Furthermore, the SCD peaks shown in Fig.1 of \ 
\cite{VossWebb} (left panel in Fig. \ref{SCDPeaks})) certainly do not
exhibit independence of temperature below $100mK$, as was claimed in the
Abstract of \ \cite{VossWebb} - the $5mK$ peak is obviously shifted compared
to the peak at $95mK$.

Also shown in Fig.\ref
{scdpeakwidths} are more recent results also claiming to vindicate the MQT
hypothesis via observations of SCD peak freezing. The lower left panel is
from \cite{Yu} with a claimed $T_{cr}=65mK$ and the lower right panel is
from \cite{Oelsner} with a claimed $T_{cr}=56mK$. Again, the approach set
out in \cite{VossWebb} was followed - temperature scales are logarithmic and
the focus is on peak widths. However, values for peak position, are
generally more precise because they are of order $\mu A$, whereas peak
widths are much smaller at $nA$. In these plots, at the lowest temperatures
where the peaks have become very narrow, the widths exhibit significant
scatter and so the added horizontal lines depicting SCD peak freezing are
misleading. Plainly the peaks do \emph{not} exhibit the temperature
independence required of MQT.

\section{A New Method for Interpreting SCD Peaks}

Gross \& Marx \cite{GrossMarx} noted that in the classical regime the width
of an SCD escape peak and also its reduced distance below the critical
current, ($1-I_{p}/I_{c}$), would both scale as $T^{2/3}$. This same result
appeared in \cite{Likharev}. Because SCD peaks become sharper at low
temperatures, their \emph{positions} ($I_{P}$) can be pinpointed easily
whereas their diminishing \emph{widths} become increasingly difficult to
extract from the peak data. \ Thus the distance test, ($1-I_{p}/I_{c}$) is
superior when looking for deviations from classical thermal activation and
this approach is now explored.

\begin{figure}
[h]
\begin{center}
\includegraphics[
height=3.2197in,
width=3.4912in
]
{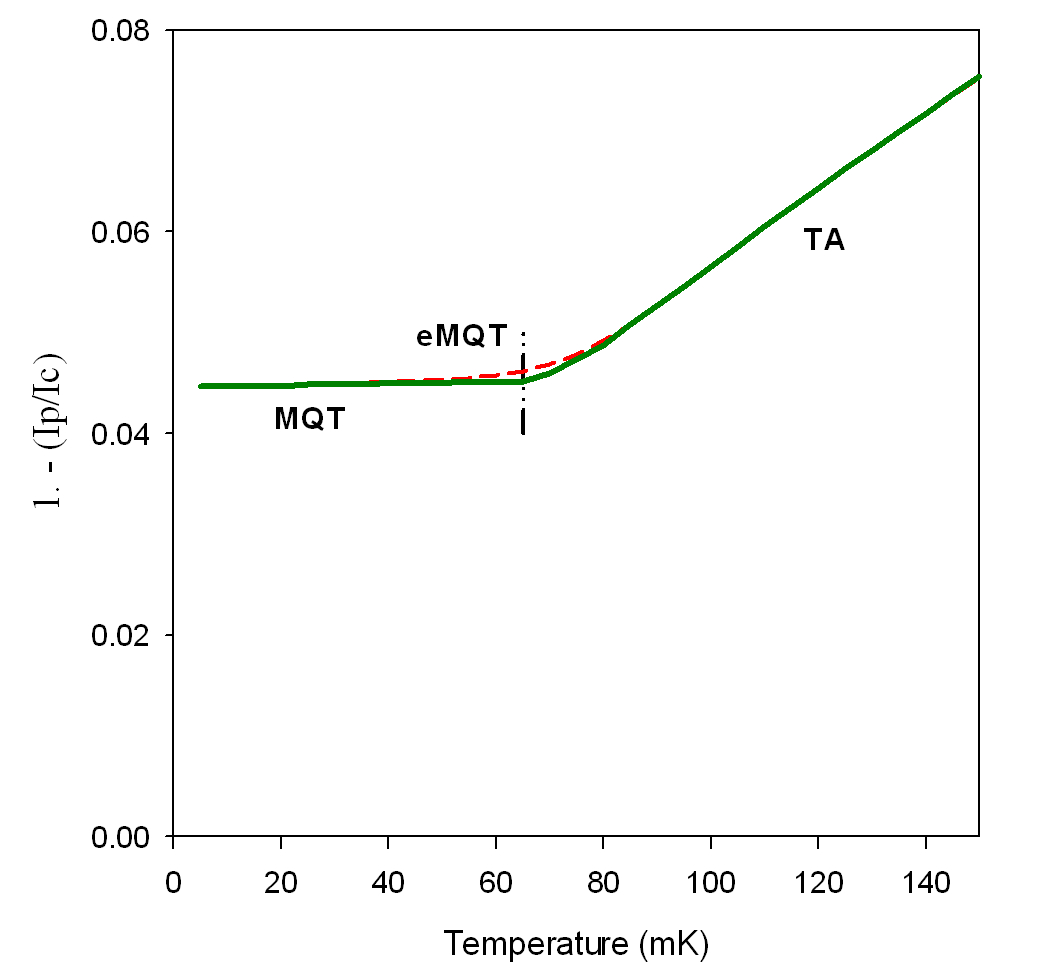}
\caption{Plot of distance of SCD peaks
from the junction critical value ($1.-I_{P}/I_{C}$) versus temperature. The
clear transition from thermal activation (TA) to the temperature independent
horizontal segment is a signature of \ the crossover to macroscopic quantum
tunneling (MQT). This feature provides a definitive test of the existence of
a macroscopic quantum state at very low temperatures. (adapted from Fig.4 in 
\protect\cite{BCJJAP} )}
\label{yudistance}
\end{center}
\end{figure}

In Fig.\ref{yudistance} we present a plot of reduced distance versus
temperature, based on the simulation described in \cite{BCJJAP}. That
simulation employed an algorithm that for the first time was able to
generate an entire SCD peak at a given temperature directly from escape
rates, either alone or in combination. \ In this example thermal activation
Eq.\ref{GAMMATA} was combined with either macroscopic quantum tunneling Eq.%
\ref{Eq18}, or with thermally enhanced macroscopic quantum tunneling Eq.\ref%
{Martinis}.\ \ It is evident in the figure that above the crossover, thermal
activation dominates, while below the crossover MQT or thermally enhanced
MQT dominate as the escape mechanism. Also apparent in the simulation output
is the fact that thermal enhancement of MQT makes little difference to the
outcome and its effects are confined to the small zone around the crossover
temperature.

The \ \ parameters for this simulation were: $I_{C}=1.957\mu A$, $C=620fF$, $%
R=300\Omega $, with a bias sweep rate of $dI/dt=0.4mA/\sec $ (based on the
experiment in \cite{Yu}). These are fairly typical values for Josephson
junctions with low damping, so apart from the exact numbers on the axes of
the plot, this characteristic shape should be seen in experimental data for
any junction - if it is undergoing a TA-to-MQT crossover.

The temperature independence that ultimately occurs below the crossover is
the signature of MQT. In \cite{Leggett2} the point is made that
\textquotedblleft for realistic parameter values the crossover should be
fairly sharp\textquotedblright ,\ and indeed that is what this simulation
shows. \ 

Any experiment must exhibit this behavior if it is to claim an observation
of MQT. \ We now apply this method to the experimental results discussed
previously.

\subsection{Experiment of Voss \& Webb}

\begin{figure}
[h]
\begin{center}
\includegraphics[
height=2.3993in,
width=3.2934in
]
{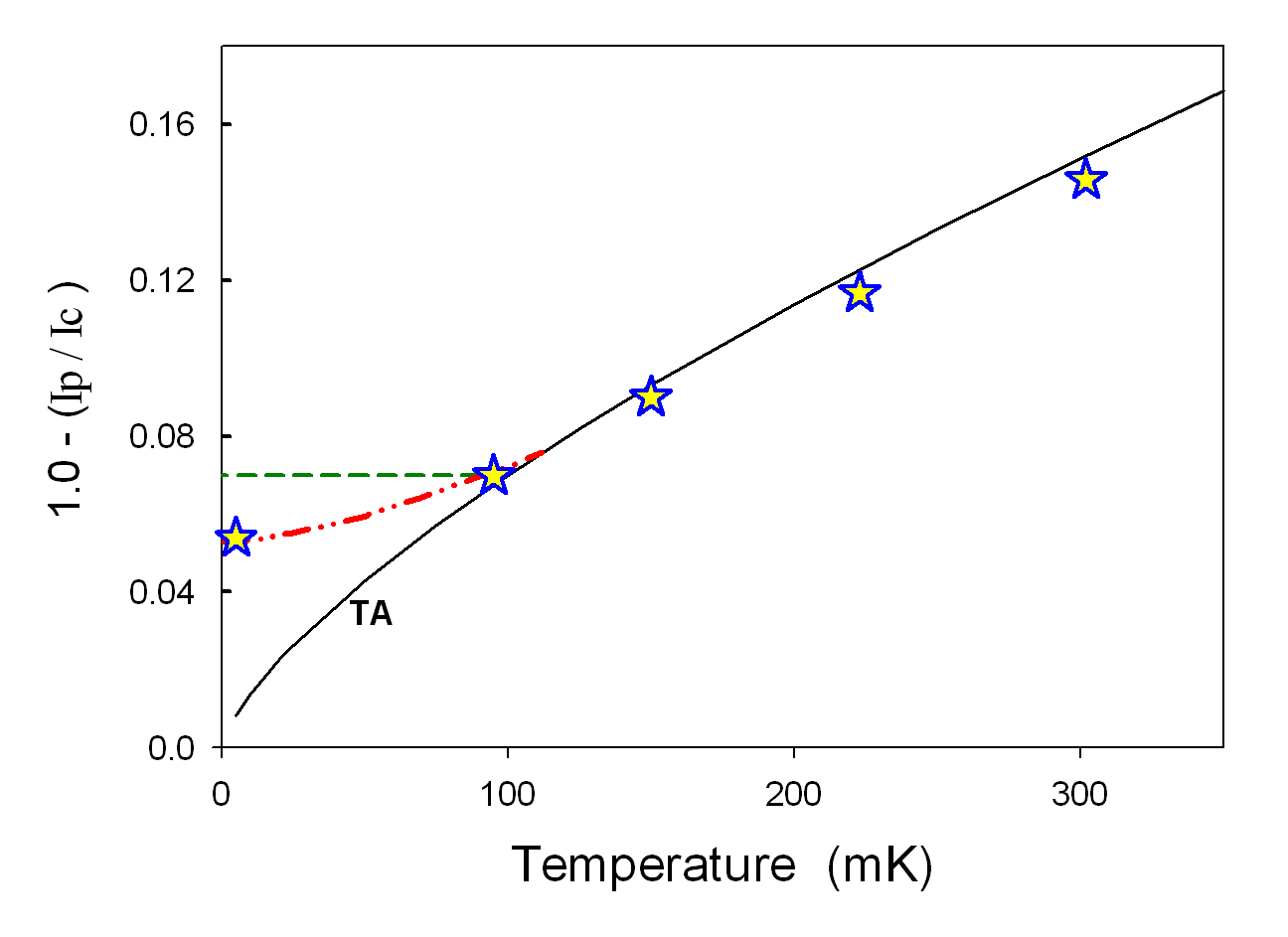}
\caption{Plot of distance of SCD peaks below the
junction critical current - data from Fig.1 in \protect\cite{VossWebb}. \
Solid line (TA) is the thermal activation characteristic. \ The dashed line
is the predicted outcome of a switch to MQT at $100mK$.}
\label{VW_distance}
\end{center}
\end{figure}

We have digitized the peak positions from Fig.1 of \cite{VossWebb}. \ The
corresponding distance parameter ($1-I_{p}/I_{c}$) is plotted as a function
of temperature, with a linear scale, in Fig.\ref{VW_distance}.

Clearly, the experimental data track the thermal activation curve down to
about $100mK$ at which point they smoothly peel away from TA. There is no
resemblance to Fig.\ref{yudistance} and no peak freezing - thus no crossover
to macroscopic tunneling.

\subsection{Experiment of Yu et al.}

\begin{figure}
[h]
\begin{center}
\includegraphics[
height=2.8312in,
width=3.7014in
]
{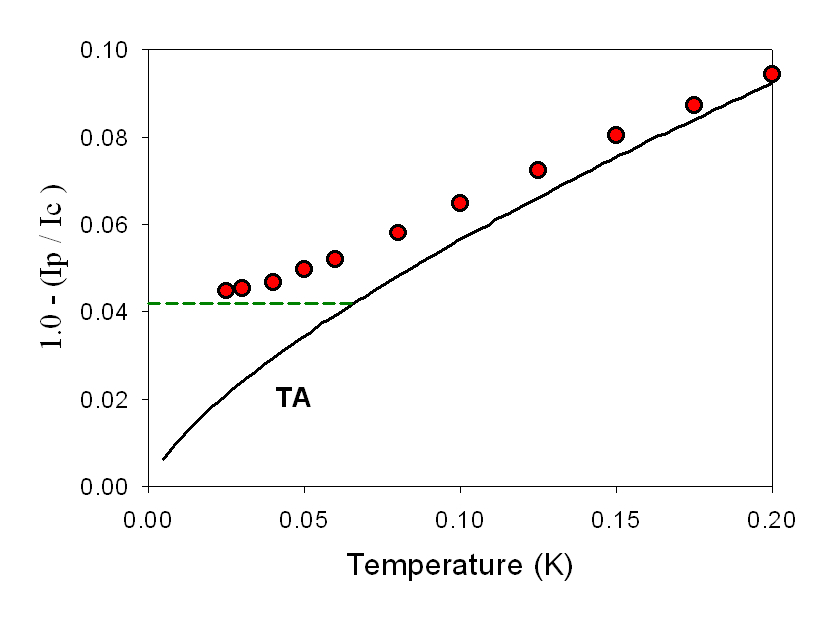}
\caption{Plot of distance of SCD peaks below the
junction critical current - data from Fig.2 in \protect\cite{Yu}. \ Solid
line (TA) is the thermal activation characteristic. \ The dashed line is the
predicted outcome of a switch to MQT.}
\label{Yu_distance}
\end{center}
\end{figure}

We have digitized the SCD peak positions from Fig.2 of \cite{Yu}. \ The
corresponding distance parameter ($1-I_{p}/I_{c}$) is plotted as a function
of temperature, with a linear scale, in Fig. \ref{Yu_distance}.

The crossover temperature for this Josephson junction was given as $65mK$.
As in the preceding case, the trend in experimental data is a gradual
peeling away from the TA characteristic. Below $65mK$ there is no
temperature independence of the peaks, as shown in Fig.\ref{yudistance}, and
hence no crossover to MQT.

\subsection{Experiment of Oelsner et al.}

\begin{figure}
[h]
\begin{center}
\includegraphics[
height=2.7992in,
width=3.7998in
]
{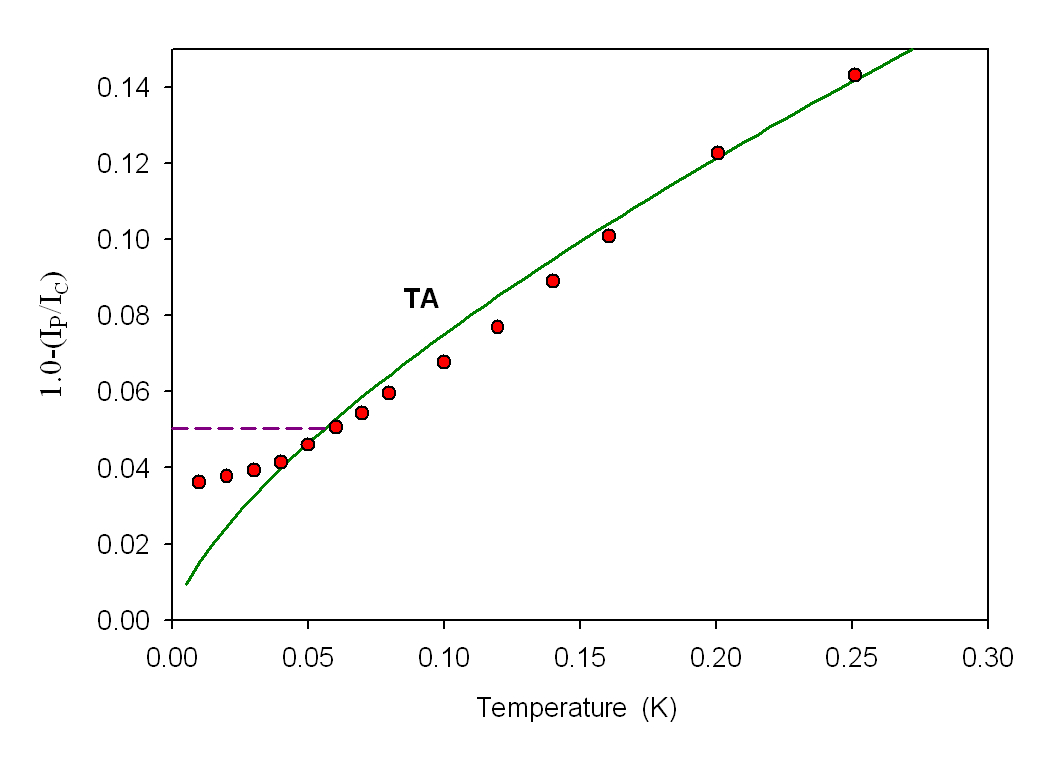}
\caption{Plot of
distance of SCD peaks below the junction critical current - data from Fig.1
in \protect\cite{Oelsner}. The reported crossover temperature was $%
T_{cr}=56mK$ and is marked with the dashed horizontal line. \ Solid curve:
the thermal activation (Kramers) characteristic.}
\label{oelsner_distance}
\end{center}
\end{figure}

We have digitized the peak positions from Fig. 1 of Oelsner et al. \cite
{Oelsner} and plot them in the distance format of ($1-I_{p}/I_{c}$) versus
temperature, see Fig. \ref{oelsner_distance}. There is obviously a
significant departure from simple thermal activation, but as in the previous
cases, there is obviously no peak freezing and therefore no evidence of a
crossover to MQT. 

\section{Josephson Qubits}

We now consider the impact of our negative conclusion regarding the
conjecture of a macroscopic quantum state for Josephson junctions, with
paricular regard to superconducting qubit architectures.

A \emph{phase qubit} is the combination of a Josephson junction and an
external source of bias current. The energy of this qubit as a function of
the junction phase $\varphi $ is

\begin{equation}
U(\varphi )=E_{J}\left[ 1-\cos \varphi \right] -E_{J}(\eta \varphi )
\label{phase}
\end{equation}%
where $E_{J}$ is the Josephson energy $\hbar I_{C}/2e$, $I_{C}$ is the
critical current of the junction, and $\eta $ is the bias current normalized
to $I_{C}$. The first term is the phase dependent junction energy and the
second term is the contribution from the bias source. This is the so-called
tilted washboard potential. Note that it is the cosine term that
superimposes potential wells on the linear bias energy.

A \emph{flux qubit} is simply a junction in a superconducting loop with an
externally applied flux bias $\Phi _{e}$. The total energy as a function of
the net flux threading the loop, $\Phi $, is 
\begin{equation}
U(\Phi )=E_{J}\left[ 1-\cos (2\pi \frac{\Phi }{\phi _{0}})\right] +\frac{1}{%
2L}\left[ \Phi -\Phi _{e}\right] ^{2}  \label{flux}
\end{equation}%
This expression utilizes a connection between junction phase $\varphi $ and
net loop flux $\Phi $ - it is $\varphi =-2\pi \Phi /\phi _{0}$ where $\phi
_{0}$ is the flux quantum. The second term is a parabola whose minimum is
positioned at a net loop flux $\Phi =\Phi _{e}$. The system energy is thus a
parabola with superimposed potential wells that originate from the Josephson
junction. With zero applied flux, the minimum of $U$ occurs at the bottom of
a single well at $\Phi =0$. With nonzero applied bias flux, the single well
at the bottom splits into two wells separated by a central barrier.

Finally, there is the \emph{charge qubit}, an arrangement basically
consisting of a Josephson junction and a very small metal island (Cooper
pair box). For this device, the total energy is%
\begin{equation}
U(n)=E_{J}\left[ 1-\cos \varphi \right] +4E_{C}\left( n-n_{g}\right) ^{2}
\label{charge}
\end{equation}%
where $E_{C}=e^{2}/2C$ is the charging energy to place a Cooper pair on the
island; $n_{g}$ is a charge, expressed in units of $2e$, induced on the
island by an applied gate voltage $V_{g}$ , and $n$ is the number of
tunneled Cooper pairs. This device architecture was modified in the
\textquotedblleft transmon\textquotedblright\ qubit, introduced in 2007,
which yielded enhanced performance leading to \textquotedblleft dramatically
improved dephasing times\textquotedblright\ \cite{Koch}.

It is readily apparent from Eqs. (\ref{phase},\ref{flux},\ref{charge}) that
all three types of superconducting qubit share a common crucial feature -
the potential wells that are created by the Josephson junctions. These wells
all have the same initial depth $E_{J}$. Discrete energy levels are the 
\textit{sine qua non} of qubits - for the case of the transmon, see for
example Fig. 3 in \cite{Koch} and Fig. 1(b) in \cite{Orlando}.

These Josephson devices can have the \emph{possibility} of being qubits only
if \ the junction itself is in a macroscopic quantum state. If the well for
a Josephson junction is not quantized even at low temperatures, then the
wells in the phase qubit, flux qubit, and charge qubit cannot be quantized
either. Then those configurations would be similar to other conventional
superconducting circuits, for example a Josephson junction coupled to a
transmission line\cite{VanDuzer}. In that case, Josephson junctions could
not form qubits.

\section{Summary}

Each of the three experiments reviewed here \cite{VossWebb} \ \cite{Yu} \ 
\cite{Oelsner} published in 1981, 2010, and 2013, claimed to confirm the
macroscopic quantum tunneling hypothesis. SCD peak freezing is a direct
consequence of macroscopic quantum tunneling, as discussed in \cite{BCJJAP}
and exemplified in the simulation results shown in Fig.\ref{yudistance}. But
such freezing did not occur in any of the three experiments. These three
devices were typical, not exceptional, so any inference must be applicable
to Josephson junctions generally. The essence of the situation is that
direct consequences of the expressions for macroscopic quantum tunneling
(Eqs.\ref{Eq18} and \ref{Martinis}) are absent from the experimental
observations. The unavoidable conclusion is that no macroscopic quantum
tunneling occurred and thus a macroscopic quantum state does not exist in
Josephson junctions.

More than a decade ago, a review of superconducting qubits \cite
{ClarkeWilhelm} stated: \textquotedblleft The first evidence of quantum
behavior in a Josephson junction came from experiments in which macroscopic
quantum tunneling was found to occur and energy levels were found to be
quantized\textquotedblright ; clearly this was a rush to judgement.

\begin{acknowledgement}
The author expresses his appreciation to Matteo Cirillo for his insightful
comments on this work. Niels Gr\o nbech-Jensen made significant
contributions to the earlier phases of this project.
\end{acknowledgement}

\end{document}